# On electrical conductivity of melts of boron and its compounds under pressure

Vladimir A. Mukhanov  and  Vladimir L. Solozhenko[*]

*LSPM–CNRS, Université Paris Nord, 93430 Villetaneuse, France*

The electrical conductivity of melts of boron and its carbide ($B_4C$), nitride (BN), and phosphide (BP) has been studied at pressures to 7.7 GPa and temperatures to 3500 K. It has been shown that these melts are good conductors with specific electrical conductivity values comparable with that of iron melt at ambient pressure.

The data on electrical conductivity of melts under pressure are extremely sparse and limited by temperatures not exceeding 1400 K [1]. In the present study an experimental evaluation of the specific electrical conductivity of melts of boron and its refractory compounds, namely, carbide ($B_4C$), nitride (BN), and phosphide (BP), at pressures to 7.7 and temperatures to 3500 K was performed for the first time.

The melting was studied in a high-pressure cell of the toroid-type high-pressure apparatus, which makes it possible to attain and keep high (to 3500 K) temperatures for 3–5 min at pressures 2–8 GPa [2]. The cell was heated by passing the direct current (to 600 A) through end electrical inputs made of pressed Ceylon graphite powder and cylindrical heater of spectral graphite. It was found that in the 2300–3000 K temperature range the total electrical resistance of cells filled with hexagonal graphite-like boron nitride ($\sigma = 5(3) \times 10^2\ \Omega^{-1}\cdot m^{-1}$) and fianite ($Zr_{0.85}Y_{0.15}O_{1.925}$) ($\sigma < 1 \times 10^3\ \Omega^{-1}\cdot m^{-1}$) is practically independent of temperature and varies from $9(1) \times 10^{-3}\ \Omega$ at 2.6 GPa to $6.5(5) \times 10^{-3}\ \Omega$ at 7.7 GPa, that is caused mainly by a decrease of the resistance of end electrical inputs with pressure increase. A sample was placed directly into a graphite heater, and at the specified pressure we measured the total electrical conductivity of the cell at different temperatures up to the melting of the substance under study. The specific electrical conductivity of the sample was calculated from Ohm's law with allowance made for the sample dimensions (height ~3 mm and diameter ~4 mm) as well as for electrical conductivity ($\sigma = 1.4 \times 10^5\ \Omega^{-1}\cdot m^{-1}$ at 2500 K and 5 GPa) and dimensions of the graphite heater, which contacts a sample. Errors of the σ determination stem from errors of the assessment of dimensions of the sample and heater in the experiment (it was suggested that they are the same for melted and quenched sample) as well as from the errors of voltage and current measurements.

A change of the total electrical conductivity of the cell starts in 1–2 s after the input of the power required for melting and ends in 10–15 s. Therefore, the effect of chemical interaction of the substances under study with a graphite heater on the melt conductivity may be neglected. According to the data of X-ray diffraction analysis of quenched samples, even in the case of elemental boron, which has the highest reactivity among the substances under study, no formation of products of interaction between melt and graphite was observed in our experiments.

---

[*] E-mail: vladimir.solozhenko@univ-paris13.fr

To test the procedure, we carried out the determination of the specific electrical conductivity of melts of high purity (99.999%) silicon and germanium at pressures from 2.6 to 7.7 GPa. The obtained values (figure) exhibit the same order of magnitude as specific electrical conductivities of melts of these elements at ambient pressure [3].

In the case of boron and its compounds an increase of a total electrical conductivity of the cell as a result of melting varies from 5.3% (B) to 11.5% ($B_4C$) at 2.6 GPa and from 5.3% ($B_4C$) to 12.7% (BP) at 7.7 GPa. The calculated values of the specific electrical conductivity of the melts are given in the figure. At high pressures the melts of boron, its carbide and phosphide are good conductors comparable with melts of silicon, germanium, and iron; and their electrical conductivity is lower than that of copper melt at ambient pressure only by an order of magnitude.

The specific electrical conductivity of boron nitride melt at 7.7 GPa and 3500 K is $2.4(12) \times 10^5 \, \Omega^{-1} \cdot m^{-1}$, which virtually coincides with conductivity of carbon melt under this pressure ($1.9 \times 10^5 \, \Omega^{-1} \cdot m^{-1}$ [6]).

Thus, over the pressure range under study the specific electrical conductivities of all studied melts are within $10^5$–$10^6 \, \Omega^{-1} \cdot m^{-1}$, which is much higher than the electrical conductivities of corresponding solids at temperatures close to the melting.

This work was supported by the Agence Nationale de la Recherche (grant ANR-2011-BS08-018) and DARPA (grant W31P4Q1210008).

REFERENCES

1. Brazhkin, V.V., Voloshin, R.N., Popova, S.V., and Umnov, A.G., Metallization of the melts of Se, S, $I_2$ under high pressure, *High Press. Res.*, 1992, vol. 10, no. 1–2, pp. 454–456.

2. Mukhanov, V.A., Sokolov, P.S., and Solozhenko, V.L., On melting of $B_4C$ boron carbide under pressure, *J. Superhard Mater.*, 2012, vol. 34, no. 3, pp. 211–213.

3. Schnyders, H.S. and Van Zytveld, J.B., Electrical resistivity and thermopower of liquid Ge and Si, *J. Phys. Condens. Matter.*, 1996, vol. 8, no. 50, pp. 10875–10883.

4. Glorieux, B., Saboungi, M.L., and Enderby, J.E., Electronic conduction in liquid boron, *Europhys. Lett.*, 2001, vol. 56, no. 1, pp. 81–85.

5. Babichev, A.P., Babushkina, N.A., Bratkovskii, A.M., et al., *Fizikal velichiny*: *Spravochnik* (Physical quantities: Reference book), Grigor'ev, I.S., Meilikhov, E.Z., Eds., Moscow: Energoatomizdat, 1991.

6. Togaya, M., Electrical property changes of liquid carbon under high pressures, *J. Phys. Conf. Ser.*, 2010, vol. 215, art. 012081.

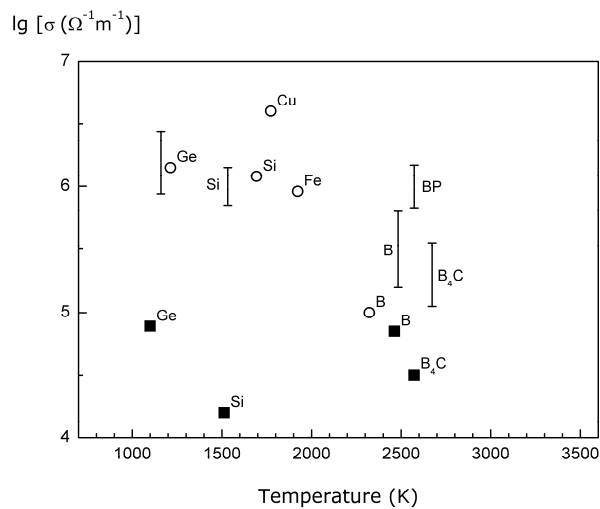 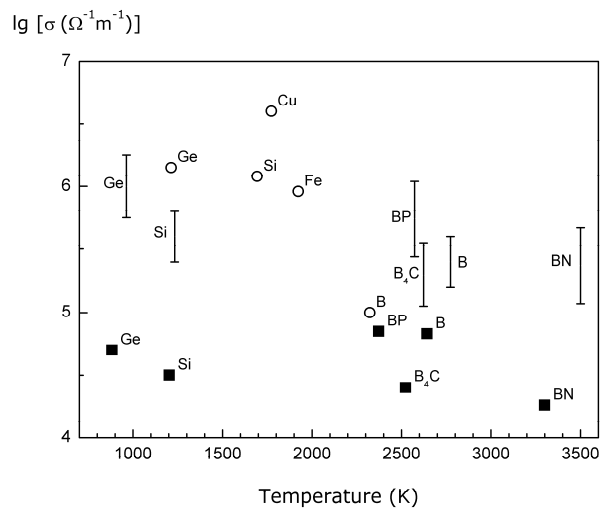

Figure  Specific electrical conductivities of melts of boron and its compounds at the melting temperature at pressures 2.6 GPa (*a*) and 7.7 GPa (*b*). For comparison, the figure shows values of the electrical conductivities of melts of some elements (open circles indicate the literature data at ambient pressure [3-5], black squares indicate the electrical conductivities of solids at temperatures close to their melting.